\documentclass[aps,prl,twocolumn,superscriptaddress,showpacs]{revtex4}
\usepackage{amsfonts}
\usepackage{amssymb}
\usepackage{amsmath}
\usepackage{graphicx}

\usepackage{dcolumn}   
\usepackage{bm}        
\usepackage{flafter}

\begin{document}

\title{Is LaO$_{1-x}$F$_x$FeAs an electron-phonon superconductor ?}
\author{L. Boeri}
\affiliation{Max-Planck-Institut f\"{u}r Festk\"{o}rperforschung,
Heisenbergstra$\rm \beta$e 1, D-70569 Stuttgart, Germany}
\author{O.V. Dolgov}
\affiliation{Max-Planck-Institut f\"{u}r Festk\"{o}rperforschung,
Heisenbergstra$\rm \beta$e 1, D-70569 Stuttgart, Germany}

\author{A.A. Golubov}
\affiliation{Faculty of Science and Technology and MESA+ Institute for Nanotechnology,
University of Twente, 7500 AE Enschede, The Netherlands}

\date{\today}

\begin{abstract}
In this paper we calculate the electron-phonon coupling of the newly-discovered
superconductor LaO$_{1-x}$F$_x$FeAs from first-principles,
using Density Functional Perturbation Theory.
For pure LaOFeAs, the calculated electron-phonon coupling constant $\lambda=0.21$
and logarithmic-averaged frequency
$\omega_{ln}=206 K$, give a maximum $T_c$ of 0.8 K, using the standard
Migdal-Eliashberg theory.
For the $F-$doped compounds, we predict even smaller coupling constants,
due to the strong suppression of the electronic Density of States at the
Fermi level.
To reproduce the experimental $T_c=26 K$, a 5-6 times larger coupling constant
would be needed.
Our results indicate that electron-phonon coupling is not sufficient to explain
superconductivity in the newly-discovered LaO$_{1-x}$F$_x$FeAs superconductor,
probably due to the importance of strong correlation effects.
\end{abstract}

\maketitle

The very recent report of superconductivity with the
remarkable $T_c$ of  $26 K$ in La[O$_{1-x}$F$_x$]FeAs~\cite{LFAO:tc:kamihara}
has stimulated an intense experimental and theoretical activity,
aimed at identifying the possible superconducting mechanism.
This compound belongs to a family of
quaternary oxypnictides of the form LnOMPn,
where Ln=La and Pr, M=Mn, Fe, Co
and Ni; Pn=P and As, synthesized in 1995.~\cite{LFAO:syn1:zimmer}
Superconductivity was first reported in LaOFeP, with a
relatively low $T_c$ of $\sim 7$ K~\cite{LaOFeP:exp:kamihara},
and later in F-doped LaOFeAs, with a maximum T$_c$ of 26 K at x=0.12
(apparently pure LaOFeAs shows no superconductivity).

The first bulk measurements on a sample with $x=0.1$
have shown that F-doped LaOFeAs has a relatively small in-plane coherence length
($\xi_{ab}=81 \AA$) and a T-dependent Hall coefficient~\cite{LFAO:exp:zhu},
the electronic specific heat displays a vanishingly small jump at $T_c$,
and its behavior under magnetic field ~\cite{LFAO:cv:mu} as well as point-contact
spectroscopy ~\cite{LFAO:cv:wen} suggest the presence of nodes in
the superconducting gap.
All these observation suggest a strong analogy with the high-$T_c$
cuprates.

A recent LSDA calculation predicts that pure LaOFeAs is on the verge of a
ferromagnetic instability, due to a very high Density
of States (DOS) of Fe $d$ electrons at the Fermi level.
~\cite{LFAO:DFT:singh}
A DMFT calculation in the paramagnetic phase,
including  strong-correlation
effects beyond LDA, shows that
 for $U=4$ eV,
a large amount of spectral weight is shifted away
from the Fermi level, and the undoped system
has a bad metallic behaviour.\cite{LFAO:DMFT:haule}
Both papers rule out standard $e-ph$ theory as a possible explanation
for superconductivity, without estimating the magnitude of the $e-ph$
coupling constant.

In this Letter, we calculate from first-principles the electron-phonon
properties of LaOFeAs, using Density Functional Perturbation Theory~\cite{DFT,SC:savrasov:band}.
Similar calculations, in conjunction with Migdal-Eliashberg theory,
reproduced the superconducting properties of many standard
$e-ph$ superconductors~\cite{SC:savrasov:band}, including MgB$_2$~\cite{MgB2}
with considerable accuracy.
On the other hand, they fail dramatically in the
the High-T$_c$ cuprates,~\cite{HTSC:DFT:savrasov,HTSC:DFT:others} where the Local Density Approximation
is not sufficient to describe the strong local electronic correlations,
and their interaction with phonons.~\cite{eph:theory:gunnarsson}

Our calculations show that
LaOFeAs is intrinsically a very poor $e-ph$ superconductor, with a very weak
$e-ph$ coupling distributed evenly over several phonon branches.
For electron-doped LaOFeAs we calculate an upper limit for the $e-ph$
coupling constant $\lambda \sim 0.21$, which, together
with $\omega_{ln}=206 K$, is a factor 5 too small
to account for the observed $T_c=26 K$.
\begin{figure}[h!tbp]
\includegraphics*[width=6cm]{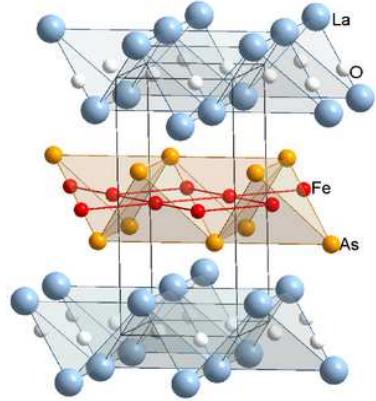}
\caption{\label{fig:fig1}(color online) Crystal structure of LaOFeAs.}
\end{figure}

LaOFeAs crystallizes in a tetragonal crystal structure (space group 129),
with $a = 4.035 (3.996 ) \AA$, $c= 8.741 (8.636) \AA$;~\cite{note:lattice}
La and As atoms occupy $2c$ Wyckoff positions,
with $z=0.1415 (0.1413)$ and $z=0.6512 (0.6415)$ respectively;
O and Fe atoms occupy $2a$ and $2b$ Wyckoff positions.

The structure, depicted in Fig.~\ref{fig:fig1}, consists of alternating
Fe-As and La-O layers.
Fe and O atoms sit at the center of slightly distorted As and La tetrahedra;
the As tetrahedra are squeezed in the $z$ direction; the Fe-As distance
is 2.41 (2.34) $\AA$, and the As-Fe-As angles are either 107.5 (105.81) or
113.5 (117.1) degrees.
Fe atoms also bond to other Fe atoms in plane, which are arranged on a square
lattice at a distance of $2 .85 (2.83) \AA$.

The gross features of the band structure of LaOFeAs are
very similar to those of LaOFeP.~\cite{LFPO:DFT:Lebegue,LFAO:DFT:singh,LFAO:DMFT:haule}.
Measuring energies from the Fermi level,
O $p$ and As $p$ states form a group of 12 bands extending from
$\sim -6$ to $-2$ eV.
La-$f$ states are found at higher energies, at $\sim 2$  eV.
\begin{figure}[h!tbp]
\includegraphics*[width=8cm]{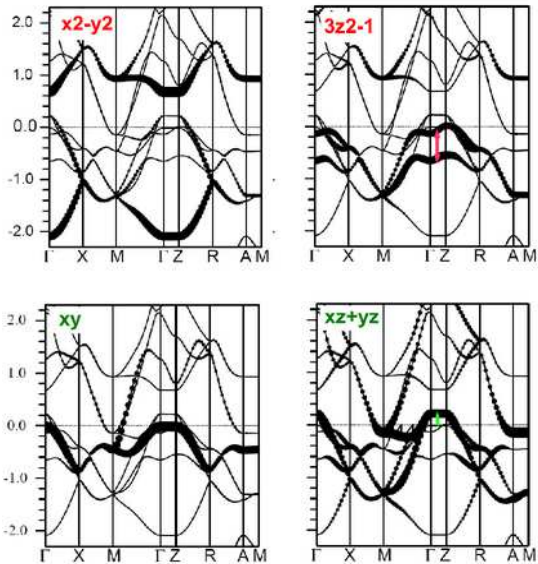}
\caption{\label{fig:fig2}(color online) Band structure of LaOFeAs, decorated with partial
characters of the $e_{g}$ (top) and $t_{2g}$ (bottom) Fe-$d$ bands.
The orientation of the coordinate system is chosen so that $Fe$-$Fe$ bonds are directed
along $x$, $y$ axes; the zero of the energy coincides with the Fermi level.
The arrows indicate the splitting induced by the elongation/shrinking
of the Fe-As tetrahedra ({\em see text}).}
\end{figure}
Apart from a weak hybridization of the $t_{2g}$ states with As p states,
at $-3$ eV, the ten Fe$-d$ states are localized in an energy window extending $\pm 2$ eV
around the Fermi level, where they give the dominant contribution
to the DOS.
The derived bands do not split simply into a lower $e_g$ ($d_{x^2-y^2}$
and $d_{3z^2-1}$) and higher $t_{2g}$ manifold, as predicted by crystal
field theory (see Fig.~\ref{fig:fig2}).
Due to the presence of Fe-Fe directed bonds,
$d_{x^2-y^2}$ orbitals, which lie along the bonds, create a pair of
 bonding-antibonding bands located at $-2$ and $+1$ eV.
 $d_{3z^2-1}$
bands  states split into two sub-bands.
Due to the distortion of the Fe-As tetrahedra,
$d_{xy}$ states are inequivalent to $d_{xz}$,
$d_{yz}$ states, and the resulting
the $t_{2g}$ bands form a complicated structure centered at $\sim -0.5$ eV.

The Fermi level cuts the band structure in a region where the
DOS is high ( 2.1 states/eV spin) and rapidly decreasing; a pseudo-gap
opens in the electronic spectrum around $0.2$ eV.
As pointed out in previous publications
 such a high DOS
at the Fermi level drives the system close to a magnetic
instability.~\cite{LFAO:DFT:singh,LFAO:DMFT:haule}

The Fermi surface comprises a doubly-degenerate cylindrical hole
pocket centered at the $\Gamma$ point, and a doubly-degenerate electron
pocket centered at the $M$ point; these sheets
have a dominant $d_{xz},d_{yz}$ character.
A small $3D$ pocket centered around the $\Gamma$ point is also present
(see Fig.3 of Ref.~\cite{LFAO:DFT:singh}).
The plasma frequencies are strongly anisotropic ($\omega_{xx}=2.30, \omega_{zz}=0.32$ eV).
The distortion (elongation or shrinking) of the Fe-As tetrahedra modulates
the splitting of the two $d_{3z^2-1}$
  bands, and the relative splitting between
$xy$ and $xz$, $yz$ bands along the $\Gamma$-Z line, as indicated by
the small arrows Fig.~\ref{fig:fig2}.
A $1 \%$ percent
 compression of the tetrahedra along the $c$ direction
changes the splitting of the two
$d_{3z^2-1}$ by $\sim 0.1 eV$, driving one of them closer
or further from the Fermi level.
This explains why the position of the $3dz^2-1$ band, and
the weight of the associated DOS, varies in literature, depending
on the crystal structure used.~\cite{LFAO:DFT:singh,LFAO:DMFT:haule}

\begin{figure}[h!tbp]
\includegraphics*[width=8cm]{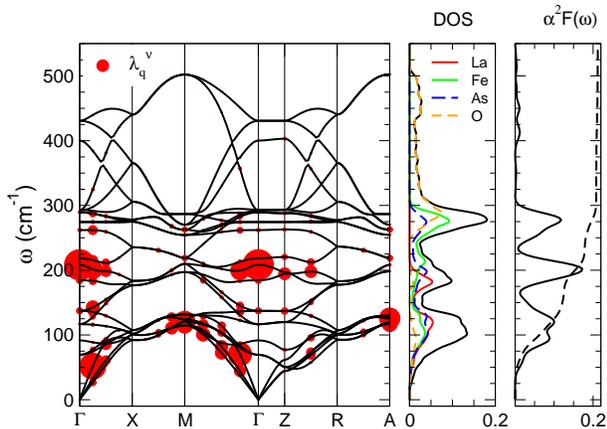}
\caption{\label{fig:fig3}
(color online) Electron-phonon properties of LaOFeAs.
{\em Left:}
Phonon dispersion relations; the radius of the symbols is proportional
to the partial $\lambda$ of each phonon mode, Eq.~(\ref{eq:eq3}).
For $\Gamma$ and $Z$ points, where the coupling diverges numerically,
we use a different scaling factor.
{\em Middle:}
Atom-projected phonon DOS. The projection on in- and out-of-plane modes
(not shown) does not show any clear separation between the patterns of vibration.
{\em Right:}
Eliashberg function $\alpha^2 F(\omega)$ (full line) and frequency-dependent
electron-phonon coupling $\lambda (\omega)$ (dashed line),
Eqs.~\ref{eq:eq1}-\ref{eq:eq2}.}
\end{figure}

Fig.~\ref{fig:fig3} summarizes
the $e-ph$ properties of LaOFeAs;
the results refer to pure LaOFeAs
in the paramagnetic phase.
It has been shown that the pure compound is
close to a magnetic instability and
to a metal-insulator transition due to electronic correlation.
~\cite{LFAO:DFT:singh,LFAO:DMFT:haule}
Electron doping strongly suppresses the tendency to magnetism and
reduces strong
correlation effects, and assuming a paramagnetic ground state
is probably appropriate for the F-doped compound.
Also, we checked by calculations in the virtual crystal approximation that
the effect of $F$ doping is well described by a rigid-band model, and the only
effect of doping is a rigid-band shift of the Fermi level,
in a region where the electronic DOS is lower (a $10 \%$ doping
correspons to a 40 $\%$ reduction of the DOS).
Therefore, the results for the undoped compound can be considered
representative also for the electron-doped compound,
provided that the reduced DOS is taken into account.

In the left panel of Fig.~\ref{fig:fig3}, we show the calculated phonon dispersion relations
of LaOFeAs; our frequencies are in very good agreement with those of Ref.~\cite{LFAO:DFT:singh},
 where a slightly different crystal structure was used.
In the middle panel of the same figure, we show the atom-projected phonon DOS.
The spectrum extends up to 500 cm$^{-1}$; the vibrations of O atoms are well
separated in energy from those of other atomic species, lying at $\omega > 300 cm^{-1}$.
The vibrations of La, Fe and As occupy the same energy
range, and the eigenvectors have a strongly mixed character.
Similarly to the electronic bands, the phonon branches have very little dispersion
in the $z$ direction.
Analyzing the evolution of the phonon eigenvectors in the Brillouin Zone (BZ)
reveals that there is no clear
separation between in and out-of-plane vibrations, as it often happens in layered compounds.
The three major peaks in the phonon DOS at $\omega = 100, 200$ and $300$ cm$^{-1}$
do not show a definite in-plane or out-of-plane character, and cannot be easily traced back to
a single vibration pattern.
This complicates the interpretation of the Eliashberg spectral function $\alpha^2 F(\omega)$,
shown in the rightmost panel of Fig.~\ref{fig:fig3}, together with the
frequency-dependent $e-ph$ coupling constant $\lambda(\omega)$:
\begin{eqnarray}
\alpha^2 F(\omega)&=&\frac{1}{N\left( 0\right) }\sum_{nm\mathbf{k}}\delta (\varepsilon _{n\mathbf{%
k}})\delta (\varepsilon _{m\mathbf{k+q}}) \times
\nonumber
\\
&\times&
\sum_{\nu \mathbf{q}}|g_{\nu ,\,n%
\mathbf{k},\,m\left( \mathbf{k+q}\right) }|^{2}\delta (\omega -\omega _{\nu
\mathbf{q}});
\label{eq:eq1}
\\
\lambda(\omega)&=&2\int_{0}^{\omega } d\Omega \alpha ^{2}F(\Omega )/\Omega
\label{eq:eq2}
\end{eqnarray}
A comparison of the Eliashberg function with the phonon DOS shows that, except for the high-lying O modes,
which show very little coupling to electrons,  the $e-ph$ coupling is evenly distributed
among all the phonon branches.
Low-frequency phonons around $100$ cm$^{-1}$ provide $\sim 75 \%$ of the total
$\lambda$, due to the $1/\Omega$ factor in Eq.~\ref{eq:eq2},
but the $e-ph$ matrix elements $g$ are comparable for all group of phonons.

It is interesting to note that this almost perfect proportionality
between the Eliashberg function and the phonon DOS is
never encountered in good $e-ph$
 superconductors, where the
coupling to electrons is usually concentrated in a few selected phonon modes.
This is best explained in terms of phonon patterns that awake dormant $e-ph$
interaction between strongly directed orbitals.

An extreme example in this sense is MgB$_2$,
which achieves a $T_c$ of 39 K thanks
to a strong coupling between bond-stretching phonons and strongly covalent
$\sigma$ bands, but the same applies also to more traditional superconductors,
such as the A15, NbC, and even normal metals.

In LaOFeAs, all phonon modes give a comparable,
small contribution to the total $\lambda$;
this indicates that there are no patterns of vibration
with a dramatic effect on
the electronic band structure around the Fermi level.
A posteriori, this is not surprising since in LaOFeAs
the only bands derived from directed bonds,
(d$_{x^2-y^2}$ in Fig.~\ref{fig:fig1}),
which could experience strong coupling to
Fe vibrations in plane,
sit far from the Fermi level.

The distribution of the coupling is also shown
in the left panel of Fig.~\ref{fig:fig3}, where the radius of the circles is proportional to
the mode $\lambda$, {\em i.e.} to the partial contribution of each phonon
mode to the total e-ph coupling
\begin{eqnarray}
\label{eq:eq3}
\lambda _{\nu \mathbf{q}}\equiv \frac{1}{\pi N\left( 0\right) }\frac{\gamma
_{\nu \mathbf{q}}}{\omega _{\nu \mathbf{q}}^{2}},
\end{eqnarray}
where $\gamma_{\nu \mathbf{q}}$ are the e-ph linewidths;
summing Eq.~(\ref{eq:eq3}) over the phonon branches $\nu$ and averaging
on the BZ give the total e-ph coupling $\lambda$.
The circles are evenly distributed over several phonon branches.
The largest couplings are concentrated around the $\Gamma (Z)$ points, where the intraband
nesting is large, and around the $M$ point, where the interband nestig between the
hole and electron cylinders take place.

The total $e-ph$ coupling constant $\lambda$, obtained by numerical integration of Eq.~(\ref{eq:eq2}) up
to $\omega=\infty$, is 0.21; this, together with a logarithmically-averaged
frquency $\omega_{ln}=205 K$, and $\mu^*=0$, gives T=0.5 K as an upper bound for T$_c$,
using the Allen-Dynes formula~\cite{Tc:allendynes:theory}.  Numerical solution of the
Eliashberg equations with the calculated $\alpha^2F(\omega)$ function gives T$_c$=0.8 K.
%
%
%
To reproduce the experimental $T_c=26 K$, a five times larger $\lambda$ would be needed, even for
$\mu^*=0$. Such a large disagreement clearly indicates that standard $e-ph$
theory cannot be applied in LaOFeAs, in line with recent theoretical works
which emphasize the role of strong electronic correlations and/or spin
fluctuations.~\cite{LFAO:DFT:singh,LFAO:DMFT:haule}

The numerical uncertainity on the calculated value of $\lambda$, connected to limited sampling of the
BZ in $\mathbf{k}$ (electrons) or  $\mathbf{q}$ (phonons) space integration, is at most 0.1, and definitely
not sufficient to raise $\lambda$ to $\sim 1.0$.
We further notice that electron doping, reducing the DOS at the Fermi level, without introducing
new bands at E$_F$, would further reduce
the value of $\lambda$.
Therefore, the value $\lambda=0.21$ for the undoped material is actually an upper bound for the value in the e-doped
compound. This value is lower than what is encountered in any known $e-ph$
superconductor; for comparison, $\lambda=0.44$ in metallic aluminum, where $T_c=1.3$ K.

In LaOFeAs, both the electronic DOS at the Fermi level, and the
value of the average phonon force constant are in line with those of other
$e-ph$ superconductors. The occurrence of a small $\lambda$
 is due to its extremely small matrix $e-ph$ elements,
connected to the strongly delocalized character of the Fe-d states at $\pm 2$
eV around the Fermi level.
This is an intrinsic property of this material, which could hardly be modified
  by external parameters, such as pressure or doping. For the same reason, our
  result is quite stable with respect to the minor differences in the
  electronic structure around the Fermi level, which have been observed in literature.

In principle,
multiband and/or anisotropic coupling could provide the missing factor 5 in the coupling,
but this is very unlikely to occur because it would require a very large anisotropy of
the distribution~\cite{Tc:dolgov:multi}. Other interactions, repulsive in the s-wave channel but
attractive in the d- or p-wave one, may increase $T_c$.

In conclusion, we have calculated the electron-phonon
properties of the newly discovered superconductor
La[O$_{1-x}$F$_x$]FeAs using Density Functional Perturbation Theory.
The undoped compound is close to a magnetic instability,
due to the presence of a very sharp peak in the electronic Density of States.
Doping with electrons moves the system away from the magnetic instability,
reducing the DOS at the Fermi level, without altering the band structure
 substantially .

Despite the high value of the DOS at the Fermi level,
the calculated value of the e-ph coupling constant for the pure compound is only $\lambda=0.21$, which is a factor
5 too small to yield the experimentally measured $T_c$ within the scope of standard ME theory.

The value $\lambda=0.21$ in LaOFeAs
is very close  to those which have been estimated using LDA
in the superconducting cuprates
($\lambda \sim 0.3$).~\cite{HTSC:DFT:savrasov,HTSC:DFT:others}
Similarly to the superconducting cuprates,
in this compound
superconductivity cannot be described using standard LDA calculations and
Migdal-Eliashberg theory.
Although our findings do not necessarily imply that superconductivity
in e-doped LaOFeAs is due to an exotic mechanism, they clearly indicate
that strong correlation effects beyond the LDA play an important role
 and must be included in any realistic description of this material.

{\bf Technical Details:}
For the atom-projected band and DOS plots in Fig.~\ref{fig:fig2} we
employed the full-potential
LAPW method~\cite{DFT:LAPW:andersen} as implemented in the
Wien2k code.~\cite{DFT:WIEN2k}
Calculations of phonon spectra, electron-phonon  coupling
and structural relaxations were performed using
planewaves and pseudopotentials with {\sc QUANTUM-espresso}~\cite{DFT:PWSCF,
note:PWSCF}.
Whenever possible, we cross-checked the results given by the two codes
and found them to be in close agreement; for consistency, we used
the same GGA-PBE exchange-correlation potential in both cases.~\cite{DFT:PBE}

{\bf Acknowledgements:}
We gratefully acknowledge O.K. Andersen for encouragement and support and J. S. Kim for useful discussion.
L.B. wishes to thank A. Toschi and G. Sangiovanni for discussion, suggestions and a critical reading of the manuscript.


\begin{thebibliography}{99}
\bibitem{note:lattice}The numbers given in parentheses refer to our fully relaxed structure.
\bibitem{note:PWSCF}
We employed ultrasoft Vanderbilt pseudopotentials,~\cite{Vanderbilt}
 with a cut-off of 40 Ryd for the wave-functions, and 320 Ryd for the charge densities.
The $\mathbf{k}$-space integration (electrons) was approximated by a summation
 over a 8 8 4 uniform grid in reciprocal space, with a Gaussian smearing
of 0.02 Ryd for self-consistent cycles and relaxations; a much finer
(16 16 8) grid was used for evaluating Densities of States (DOS) and
electron-phonon linewidths.
%
Dynamical matrices and electron-phonon linewidths were calculated
on a uniform 442 grid in $\mathbf{q}$-space; phonon dispersions and DOS
were then obtained by Fourier interpolation of the dynamical matrices,
and the Eliashberg function by summing over individual linewidths and phonons.
%
To check the effect of nesting on the e-ph linewidhts, we also calculated
selected $\mathbf{q}$ points on a 882 grid, shown in Fig.~\ref{fig:fig3}.
\end{thebibliography}
\end{document}